\documentclass[10, twocolumn, nofootinbib]{revtex4-1}
\usepackage[margin=1in]{geometry}
\usepackage[parfill]{parskip}
\usepackage{graphicx}
\usepackage{amssymb}
\usepackage{epstopdf}
\usepackage{amsmath}
\usepackage{hyperref}
\usepackage{multirow}
\usepackage{caption}
\usepackage{subcaption}
\usepackage{float}
\usepackage{titlesec}
\usepackage[utf8x]{inputenc}

\renewcommand{\vec}{\mathbf}
\newcommand{\be}{\begin{equation}}
\newcommand{\ee}{\end{equation}}

\newcommand{\bea}{\begin{eqnarray}}
\newcommand{\eea}{\end{eqnarray}}

\newcommand{\bci}{\begin{compactitem}}
\newcommand{\eci}{\end{compactitem}}

\newcommand{\fr}{\frac}

\def\le{\left}
\newcommand{\ri}{\right}
\newcommand{\eq}{\equiv}
\newcommand{\rbao}{r_{BAO}}
\def\a{\alpha}
\newcommand{\Ode}{\Omega_{DE}}
\newcommand{\Om}{\Omega_m}
\newcommand{\Or}{\Omega_{r}}

\newcommand{\rde}{\rho_{DE}}

\newcommand{\La}{\Lambda}

\begin{document}

\title{Probing a Steep EoS for Dark Energy with latest observations}
\author{Mariana Jaber and Axel de la Macorra }

\affiliation{Instituto de Fisica, Universidad Nacional Autonoma de Mexico,\\  A.P. 20-364, 01000, Mexico D.F., Mexico}

\begin{abstract}
	We present a parametrization for the Dark Energy Equation of State ``EoS"  which has a rich structure, performing a transition at pivotal redshift $z_T$ between the  present day value $w_0$ to  an early time $w_i=w_a+w_0\equiv w(z\gg0)$ with a steepness given in terms of $q$ parameter. The proposed parametrization is  $w=w_0+w_a(z/z_T)^q/(1+(z/z_T))^q$, with $w_0$, $w_i$, $q $ and $z_T$ constant parameters. It reduces to  the widely used EoS $w=w_0+w_a(1-a)$ for $z_T=q=1$. This transition is motivated by scalar field  dynamics such as for example quintessence models.  We study if a late time transition is favored by BAO measurements combined with local determination of $H_0$ and information from the CMB. According to our results, an EoS with a present value of $w_0 = -0.92$ and a high redshift value $w_i =-0.99$, featuring a transition at  $z_T =  0.28$ with an exponent $q = 9.97$ was favored by data coming from local dynamics of the Universe (BAO combined with $H_0$ determination). 
We find that a dynamical DE model allows to  simultaneously  fit   $H_0$ from  local determinations  and Planck CMB measurements, alleviating the tension obtained in a $\Lambda$CDM model.
	Additionally to this analysis we solved numerically the evolution of matter over-densities in the presence of dark energy both at background level and when its perturbations were considered. We show that the presence of a steep transition in the DE EoS gets imprinted into the evolution of matter overdensities and that the addition of an effective sound speed term does not erase such feature.

\end{abstract}

\maketitle

\section{INTRODUCTION}
 
We live in a  particular epoch of the cosmic history characterized by the acceleration in the expansion rate of the Universe.
Although its cause is unknown this acceleration is described  as the consequence of a Cosmological Constant, $\Lambda$,  with  density  $\rho_{\La}$ constant in space and time. 
Despite its simplicity, there is no fundamental understanding of its origin and this framework has serious theoretical issues namely the coincidence  and fine-tunning problems(\cite{RevModPhys.61.1,Sahni:2002kh}). 
For this reason alternative models that either modify gravity  at large scales as prescribed by General Relativity or introduce a dynamical Dark Energy (DE) component have arisen. 
Dynamical dark energy models are often characterized by the DE equation of state (EoS), $w \equiv P/\rho$, which is the ratio of the DE pressure to its density.
Since DE properties are still unknown, several models to parametrize its EoS as a function of time, $w(z)$,  have arisen in the literature (\cite{Doran:2006kp,  Rubin:2008wq, 2009ApJ...703.1374S, 2010PhRvD..81f3007M, Hannestad:2004cb, Jassal:2004ej, Ma:2011nc, Huterer:2000mj, Weller:2001gf, Huang:2010zra, delaMacorra:2015aqf}). 
One of the most po\-pu\-lar among them is the CPL parametrization (\cite{Chevallier:2000qy},\cite{Linder:2002et}), widely used in cosmological observational analysis. 
The present value of DE EoS is restricted by observations to be close to $-1$  ($ w = -1.019^{+0.075}_{-0.080}$ according to the 95$\%$ limits imposed by \emph{Planck } data combined with  other astrophysical measurements  \cite{Ade:2015xua}).
Nevertheless, the DE behavior and its properties at different cosmic epochs are much poorly constrained by current cosmological observations. 
According to astrophysical observations our Universe is flat and dominated at present time by  the DE component (\cite{Ade:2015xua}), so data coming from late-time, low-redshift measurements such as Baryon Acoustic Oscillations (BAO) from Large Scale Structure surveys 
are those best suited for its analysis.

The aim of this work is to determine the late time dynamics of DE  through its EoS, and in particular we are interested in studying if a transition in $w(z)$ takes place.
To that end the  parametrization used  is $w(z) = w_0 + w_a(z/z_T)^q/[1+(z/z_T)]^q$, with $w_0$, $w_a=w_i-w_0$, $q$, and $z_T$  constant parameters. This EoS a\-llows for a steep transition for a large value of $q$ at the  pivotal point $z_T$, which is prompted by scalar field  dynamics such as quintessence models \cite{delaMacorra:1999ff} and motivated in \cite{delaMacorra:2015aqf}, where a new parametrization that  captures the dynamics of DE is presented.

The scientific community is devoting a large amount of time and resources in the quest to understand the dynamics and nature of DE, working on current (SDSS-IV \cite{Dawson:2015wdb}, DES \cite{Abbott:2005bi}) and future  (DESI \cite{Levi:2013gra, Aghamousa:2016zmz, Aghamousa:2016sne}, Euclid \cite{2011arXiv1110.3193L}, LSST \cite{2009arXiv0912.0201L}) experiments to study with  very high precision the expansion history of the Universe  and thus be able to test interesting models beyond a Cosmological Constant or Taylor expansions of the EoS of DE.

This article is organized as follows: we introduce our theoretical framework and the data sets used  in Section \ref{sect:Method}, Section \ref{sec:Background} details the analysis performed at Background level and the results obtained, Section \ref{sec:perturbs} discusses the analysis at perturbative level  while Section \ref{sec:Conclusions} summarizes our Conclusions. 
 
\section{METHOD AND DATA}
	\label{sect:Method}

\begin{figure}[t]
\centering
\includegraphics[width=\linewidth]{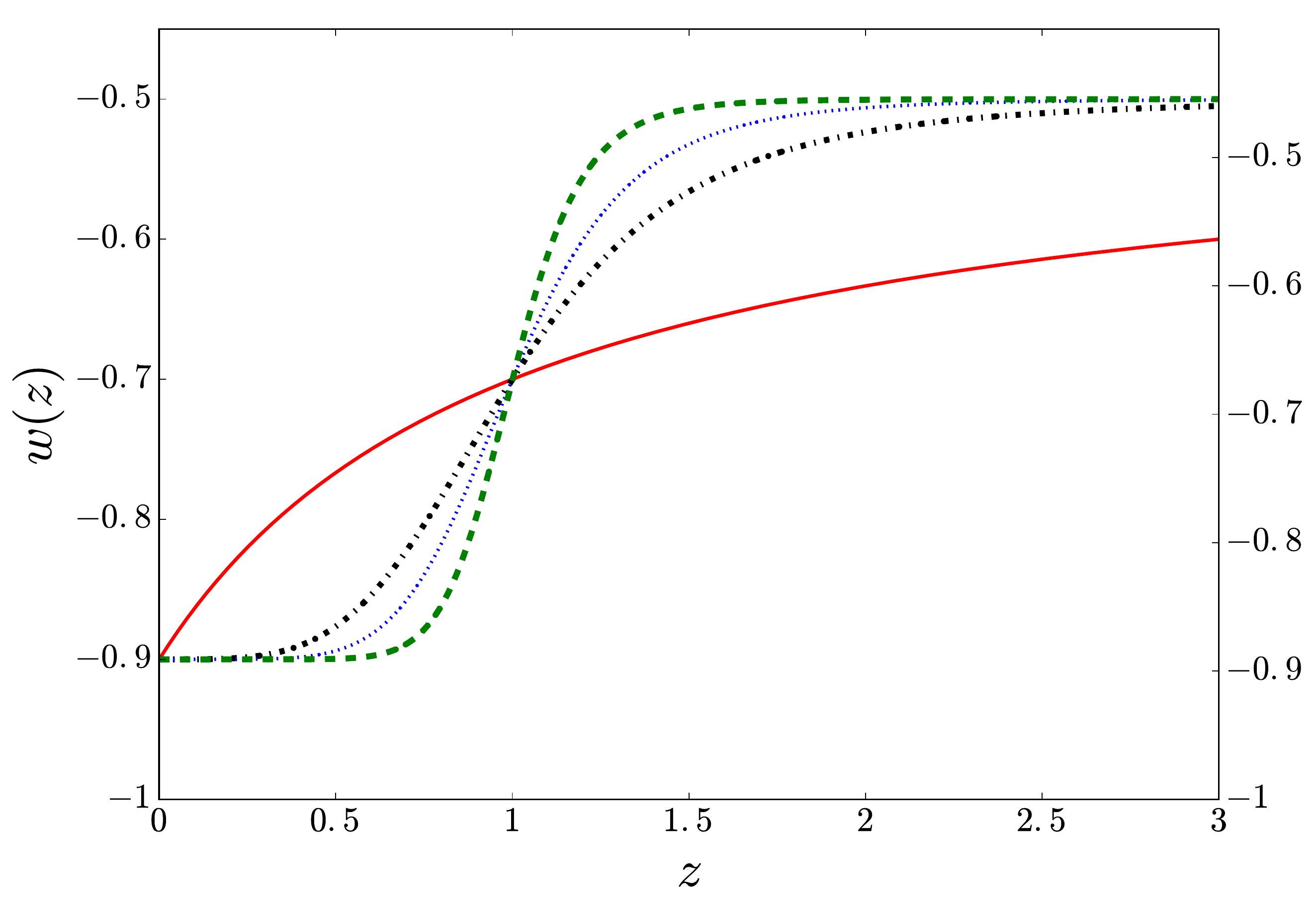}
\caption{Evolution of $w(z)$ from equation \eqref{eq:eos} with $q=1$ (solid red), $q = 4$ (black dot-dashed), $q=6$ (blue dotted), and $q = 10$ (green dashed). The other parameters were fixed to $w_0 = -0.9$, $w_i = -0.5$ and $z_T=1$. The solid red curve takes the special case $q=z_T=1$, representing the CPL parametrization \eqref{eq:CPL}.}
\label{fig:steep-eos}
\end{figure}

Within the General Relativity framework for a flat Universe and a FLRW metric we have
\be
	\label{eq:Hz}
	H(z) = H_0\sqrt{\Or(1+z)^4+\Om(1+z)^3+\Ode F(z)}
\ee
where $H\eq(\fr{da}{dt})(\fr{1}{a})$ is the Hubble parameter, $a=(1+z)^{-1}$ the scale factor of the Universe and  \ $H_0 = 100\cdot h$  is the Hubble parameter at redshift zero in units  of $km\cdot s^{-1} Mpc^{-1}$. 
The present value of matter, radiation, and DE fractional densities are given by $\Om$, $\Or$, $\Ode$, respectively. 
 The function $F(z) \eq \fr{\rde(z)}{\rde(0)}$ in equation  \eqref{eq:Hz} encodes the evolution of  DE component in terms of its EoS, $w(z)$, according to
\be
 	F(z) = exp\le( -3 \int_0^z dz'\fr{1+w(z')}{1+z'}\ri)
	\label{eq:DE-f}
\ee
where $w(z)$ specifies the evolution of the DE fluid and accordingly, the  Universe expansion rate at late times, following the dynamics set by equation \eqref{eq:Hz}. 

%
\be
w(z) = w_0 + w_a\fr{(\fr{z}{z_T})^q}{1+(\fr{z}{z_T})^q} 
\label{eq:eos}
\ee

with $w_a=w_i-w_0$  where $w_i$ and $w_0$ represent the value for $w(z)$ at large redshifts and at present day, respectively, whereas the term $ f(z)\equiv\fr{(\fr{z}{z_T})^q}{1+(\fr{z}{z_T})^q}$ modulates the dynamics of this parametrization in between both values, and takes the values $f(z=0)=0$, $f(z\rightarrow \infty)=1$ and $f(z=z_T)=1/2$, \emph{i.e.} $0\leq f(z) \leq 1$.
This EoS makes a transition between the two regimes: $w(z=0) \rightarrow w_0$, $w(z\gg0) \rightarrow w_i$, at redshift $z = z_T$, taking a value of $w(z_T) =(w_0+w_i)/2$. The parameter $q$ modulates the steepness of the transition featured: a larger value for $q$ has a steeper transition, as figure \ref{fig:steep-eos} shows. 
%
%

For $q=z_T=1$, equation \eqref{eq:eos} includes the well known CPL pa\-ra\-me\-tri\-za\-tion (\cite{Chevallier:2000qy, Linder:2002et}) as a particular case  but  it allows for a richer physical behavior. CPL EoS written in terms of scale factor reads:
\be
	\label{eq:CPL}
	w(a) =  w_0 + w_a(1-a)
\ee
from where we see that  its slope is constant and with a value $dw(a)/da=-w_a=-(w_i-w_0)$, meaning that the late time dynamics of DE is fixed from the present and initial values of the EoS. 

Clearly, taking $w_0 = w_i = -1$ in \eqref{eq:eos}, the Cosmological Cons\-tant solution, $w_{\La}\eq-1$, is recovered. 

\subsection{BAO as a cosmological probe}
\label{subsect:BAO}

Ever since its first detection (Cole et al. 2005 \cite{Colless:2003wz}, Eisenstein et al 2005, \cite{Eisenstein:2005su}) the Baryon Acoustic Osci\-lla\-tion feature has been widely used as a po\-wer\-ful tool for cosmology becoming the standard ruler of choice. 
 It has become the best way to probe late time dynamics of the Universe and in consequence that of  DE.  For that reason it is the cosmological tool  used by several experiments like 6dF \cite{Beutler:2011hx}, WiggleZ  \cite{Kazin:2014qga}, SDSS-III \cite{Anderson:2013oza},(and most recently \cite{Alam:2016hwk}),  SDSS-IV \cite{Dawson:2015wdb} and Dark Energy Survey (DES) \cite{Abbott:2005bi} and the main  probe to be implemented in future experiments like the Dark Energy Spectroscopic Instrument (DESI) \cite{Levi:2013gra, Aghamousa:2016zmz, Aghamousa:2016sne} and Euclid \cite{2011arXiv1110.3193L}.

The co\-rres\-pon\-ding size, $\rbao(z)$, is obtained by performing a spherical average of the galaxy distribution both along and across the line of sight (Bassett and  Hlozek 2010 \cite{Bassett:2009mm}):
\be
	\label{eq:rbao}
	\rbao(z) \eq \frac{r_s(z_d)}{D_V(z)}
\ee
The comoving sound horizon at the baryon drag epoch is represented by $r_s(z_d)$  and the dilation scale, $D_V(z)$, contains information about the cos\-mo\-lo\-gy used in $H(z)$:
\bea
	\label{eq:sd}
	r_s(z_d) & \eq &  \int_{z_d}^{\infty} \fr{dz}{ H(z)  \sqrt{3(R(z)+1)}}, \\
	\label{eq:DV}
	D_V(z) & \eq & \left[\fr{z (1+z)^2}{H(z)} D_A(z)^2 \right]^{1/3},
\eea
where
\be
	\label{eq:daz}
	D_A(z)=\frac{1}{1+z} \int_0^z \fr{dz'}{H(z')}
\ee

In this way, the BAO standard ruler which is set by a particular size in the spatial distribution of matter, can be used to constrain the  parameters in equation  \eqref{eq:eos}. 

While the sound horizon, $r_s(z_d)$, depends upon the physics prior to the recombination era,  given by  $z_d\approx1059$   \cite{Ade:2015xua}  and the baryon to photon ratio,  
 $R(z)\eq\frac{3\Omega_{\gamma}(z)}{4\Omega_b(z)}$,  the dilation scale, $D_V(z)$,  is sensitive to the physics of much lower redshifts, particularly to those probed  by Large Scale Structure experiments.

In this work we make use of the observational points from the six-degree-field galaxy survey (6dFGS \cite{Beutler:2011hx}), Sloan Digital Sky Survey Data Release 7 (SDSS DR7 \cite{Ross:2014qpa}) and the reconstructed value (SDSS(R)  \cite{Padmanabhan2pc}), as well as the latest result from the complete BOSS sample SDSS DR12 (\cite{Alam:2016hwk}), and the Lymann-$\a$ Forest (Ly$\a$-F) measurements from the Baryon Oscillation Spectroscopic Data Release 11 (BOSS DR11 \cite{Font-Ribera:2013wce}, \cite{Delubac:2014aqe}).  
Table  \ref{table:datarbao} summarizes them all.  
Since the volume surveyed by BOSS and WiggleZ \cite{Kazin:2014qga} partially overlap we do not  use data from the latter in this work (see details in \cite{Beutler:2015tla}).
\subsection{Local value of the Hubble Constant}
\label{subsec:h0}

The present value of Hubble constant has been determined observationally from direct measurement of the local dynamics, as in the latest work of A. Riess \emph{et al} in \cite{localhubble}, but also  from BAO measurements either  from galaxy surveys or from the Lyman-$\alpha$ forest and it can be derived as well from CMB  experiments such as Planck. 

Regarding the work of A. Riess \emph{et al}  (AR16), their best estimate in units of $km\cdot s^{-1}Mpc^{-1}$ reports a value of
  \be
  H_0=73.21\pm 1.74,
  \label{H0}
  \ee
the accuracy of which was achieved in great deal due to the utilization of maser system in NGC1258 both to calibrate and as an independent anchor for the cosmic distance ladder.

\subsection{Cosmic Microwave Background}
\label{subsec:cmb}

The Cosmic Microwave Background is the most precise cosmological data set. The angle subtended by the first peak is determined with exquisite precision  (\cite{Ade:2015xua}):
\be
\label{eq:thetastar}
\theta_*=1.04077\pm0.00032\times10^{-2}
\ee

Following  the latest report of the Planck collaboration (P15)
\cite{planck15DE} and \cite{galli2014} we use \emph{Planck TT+TE+EE+lowP } which denotes the combination of  likelihood at $l\leq30$ using TT, TE, and EE spectra with the low-$l$ temperature+polarization likelihood. 

However, it has been shown (\cite{wangmukherjee}, \cite{mukherjee}, \cite{planck15DE}) that the information of CMB power spectra can be compressed within few observables such as the angular scale of sound horizon at last scattering, $l_A\equiv\pi/\theta_*$, and the scaled distance to last scattering surface, $R \equiv \sqrt{\Omega_MH_0^2}d_{A}(z_*)$.

We keep the flat geometry and the baryon density fixed and thus we can add the CMB information to  BAO  and $H_0$ measurements by means of the observables $\{\theta_*, \omega_c\equiv\Omega_ch^2 \}$. The corresponding covariance matrix is
\begin{equation}
\label{eq:covmatcmb}
\mathcal{C}_{CMB}  =  
\bordermatrix
{~ & \omega_c & \theta_* \cr
	\omega_c & -23.5248 & -2.2078 \cr
	\theta_* & -2.207815 & 1.063561 \cr}
\times 10^{-7} 
\end{equation}
The angle of horizon at last scattering is defined to be
\begin{equation}
\label{eq:thetadec}
\theta_* \equiv \frac{r_s(z_*)}{d_A(z_*)}
\end{equation}
where $r_s(z_*)$ is the horizon size at the decoupling epoch ($z_*\approx1090.06$ according to Planck \cite{Ade:2015xua}), defined by the integral in equation \eqref{eq:sd} evaluated from $z_*$ to $\infty$, and $d_A(z_*)$ is the comoving distance to last scattering surface:
\be
\label{eq:dacmb}
d_A(z_*) = \int_{0}^{z_*}\fr{dz'}{H(z')}
\ee 

The reported value for the Hubble constant by P15 is $H_0=67.8 \pm 0.9$ \cite{Ade:2015xua}, which assumes a $\La$CDM universe and  is known to be in tension with AR16 at the $3.4\sigma$ level.

\section{BACKGROUND EVOLUTION}
\label{sec:Background}

Additionally to parameters in  equation \eqref{eq:eos} we also investigate the cons\-traints on the physical density of cold dark matter  $\omega_c \equiv \Omega_c h^2 $  and $H_0$ (or equivalently $h$),  resulting in  the set   
$\vec{\alpha}=\{w_0,  w_i,  z_T,  q,  \omega_c,  h \}$ 
and consider uniform priors on these: $h\in$ [0.5, 1], $\omega_c\in$ [0.001, 0.99], $w_0\in$[-1, 0], $w_i\in$ [-1, 0], $q\in$ [1, 10] and $z_T\in$ [0, 3].
To determine the  best-fitting values (BFV), we minimize the $\chi^2$ goodness-of-fit estimator, 
\be
\label{eq:chi2}
\chi^2=(\vec{m}-\vec{d})^T\vec{C}^{-1}(\vec{m}-\vec{d})
\ee
where $\vec{m}$ are  theoretical values for each observable (namely  $\rbao(z)$, $H_0$,  $\omega_c$,  $\theta_*$) and $\vec{d}$ the data. The corresponding covariance matrix is represented by $\vec{C}$. 

The joint analysis of the different data sets is done by adding their respective $\chi^2$ functions. Further details can be found in the Appendix \ref{appendix}.

The reported value of $\Omega_{m}$, $\rho_{DE}\equiv\Omega_{DE}h^2$ and $\rho_{\La}\equiv\Omega_{\La}h^2$ in tables summarizing our results was obtained by taking the BFV for $\omega_c$ and $H_0$ for each model, as well as the fixed value $\omega_b\equiv\Omega_bh^2=0.02225$ from P15 \cite{Ade:2015xua}.

\begin{table}
	\begin{center}
		\begin{tabular}{|c|c|c|}
		\hline
\bf{Data set} & \bf{Redshift} &$\rbao(z)$ \\ \hline
6dF  & 0.106 \cite{Beutler:2011hx} & 0.336 $\pm$ 0.015 \\ \hline
SDSS DR7  & 0.15 \cite{Ross:2014qpa} &  0.2239 $\pm$ 0.0084 \\ \hline
SDSS(R) DR7  & 0.35 \cite{Padmanabhan2pc} &  0.1137 $\pm$ 0.0021 \\ \hline

\multirow{2}{*}{SDSS-III DR12} & 0.38 \cite{Alam:2016hwk} & 0.100 $ \pm$ 0.0011\\ & 0.61 \cite{Alam:2016hwk} & 0.0691 $\pm$ 0.0007  \\ \hline

\multirow{2}{*}{SDSS-III DR11}& 2.34 \cite{Delubac:2014aqe} & 0.0320 $\pm$ 0.0013 \\ & 2.36  \cite{Font-Ribera:2013wce} & 0.0329 $\pm$ 0.0009 \\ \hline
		\end{tabular}
		\caption{$\rbao(z)$ measurements used in this work. The ones corresponding to SDSS data were inverted from the published values of $D_V(z)/s_d$ and those corresponding to Ly$\a$-F data were obtained from the reported quantities $D_A(z)/s_d$ and $D_H(z)/s_d$. }
	\label{table:datarbao}
	\end{center}
\end{table}

\begin{figure}[t]
	\centering
		\includegraphics[width=\linewidth]{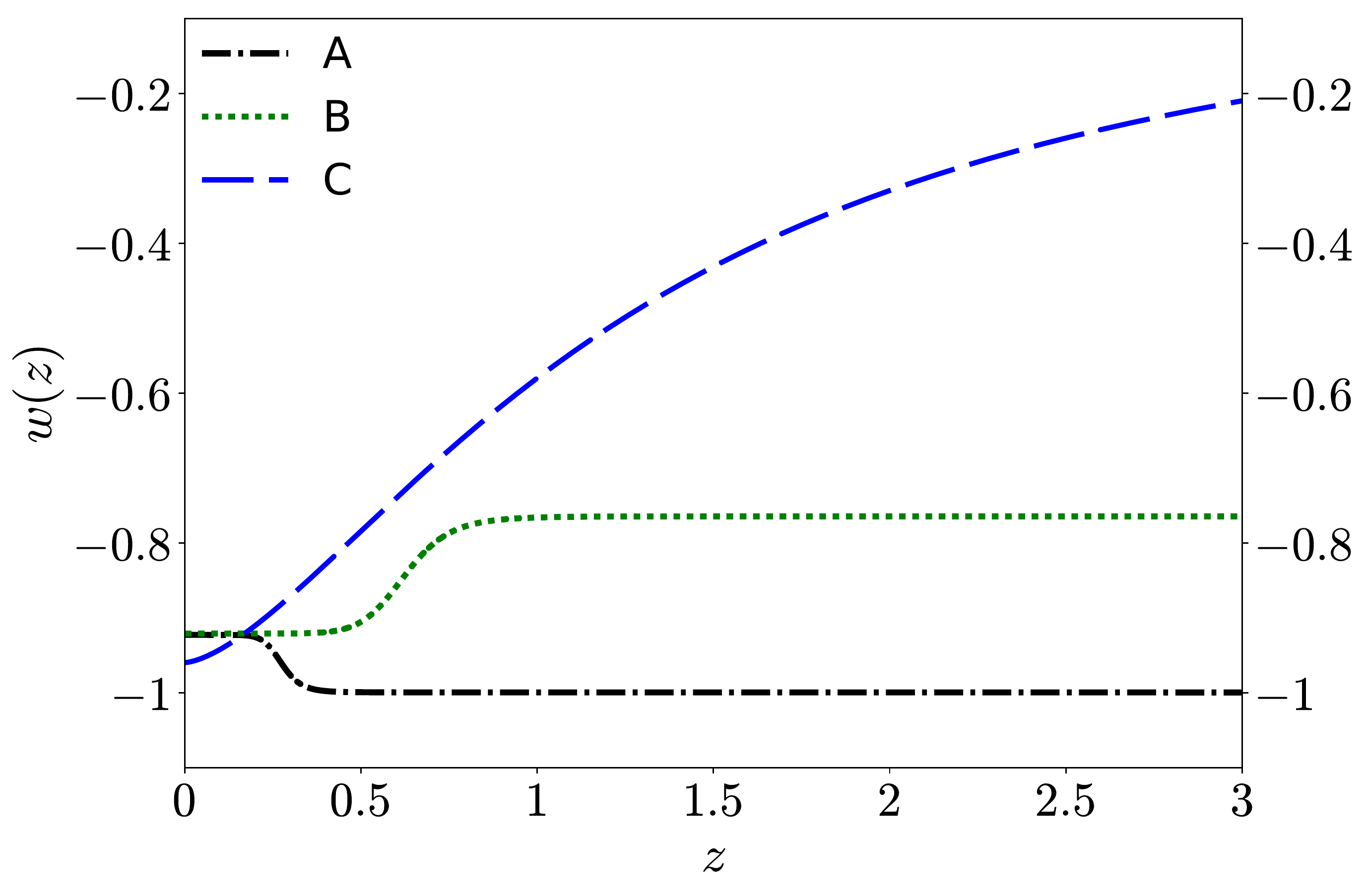}
		\caption{Evolution of the EoS $w(z)$ in equation \eqref{eq:eos} according to the best fit values reported in Table \ref{table:results}.}
		\label{fig:bfveos} 
\end{figure}

\begin{figure*}
	\centering
	\begin{subfigure}[t]{0.3\textwidth}	
		\captionsetup{width=5cm}
		\includegraphics[height=\textwidth]{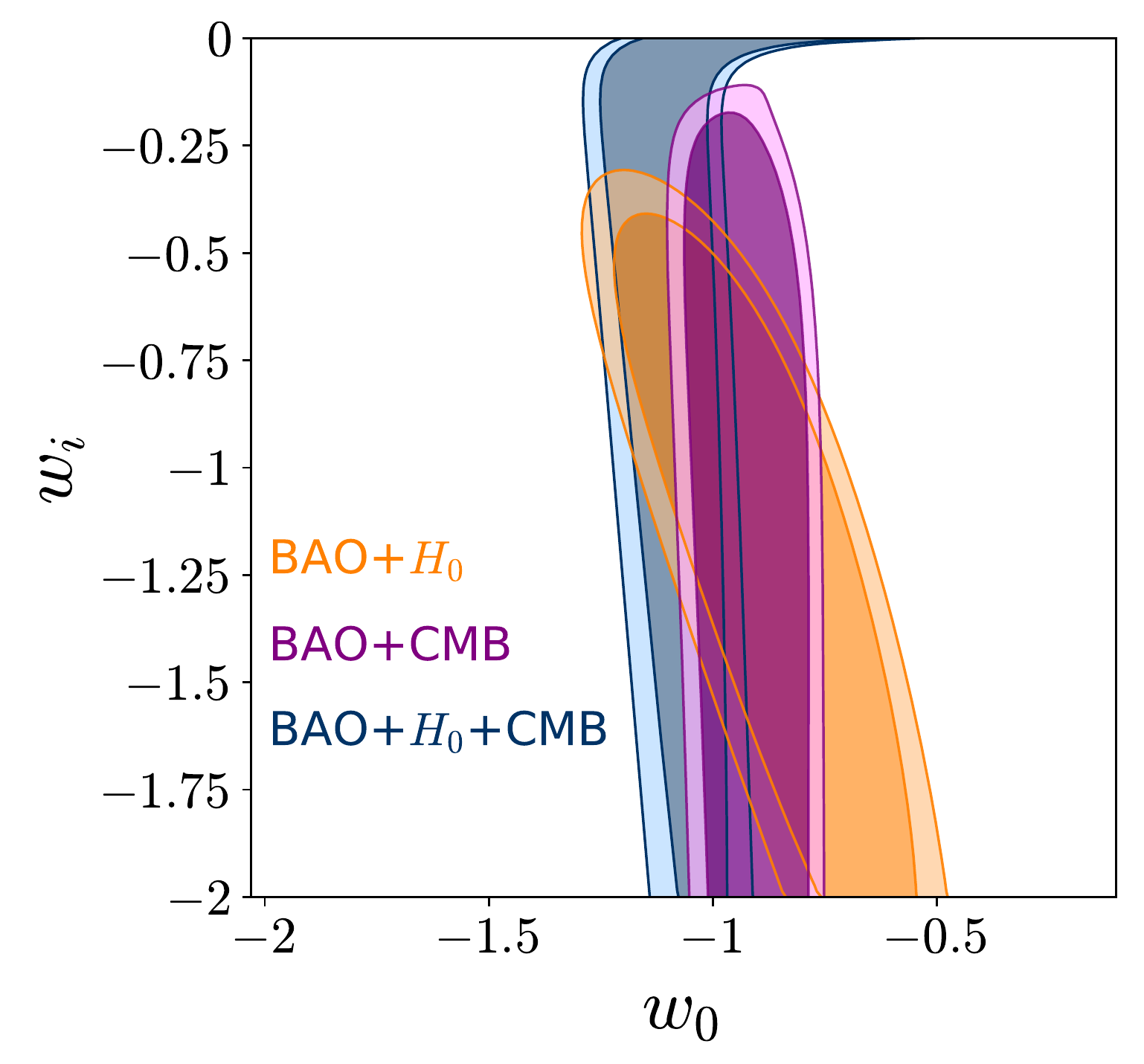}
	\end{subfigure}\quad
	\begin{subfigure}[t]{0.3\textwidth}
		\captionsetup{width=5cm}
		\includegraphics[height=\textwidth]{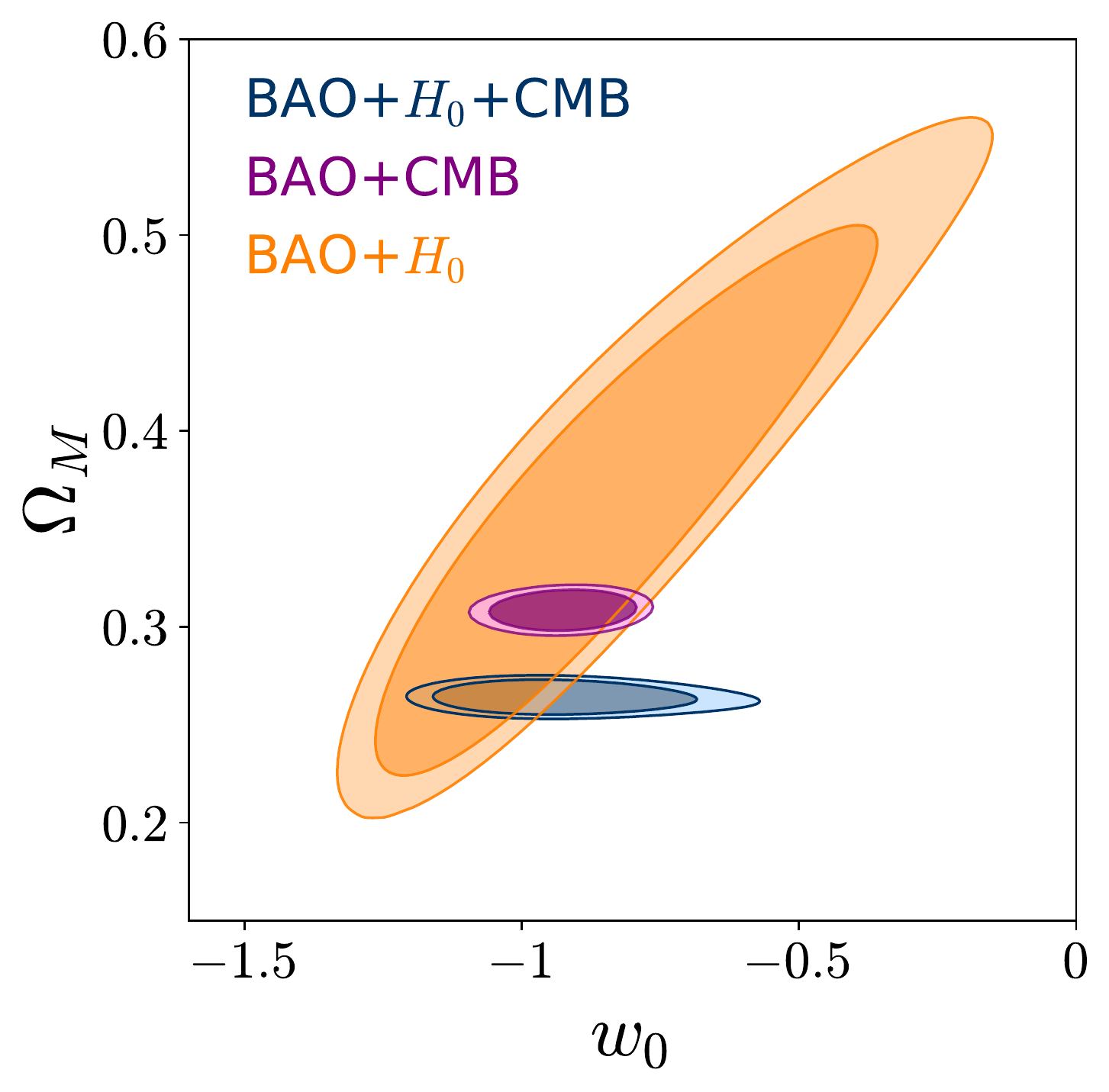}
	\end{subfigure}\quad
	\begin{subfigure}[t]{0.3\textwidth}
		\captionsetup{width=5cm}
		\includegraphics[height=\textwidth]{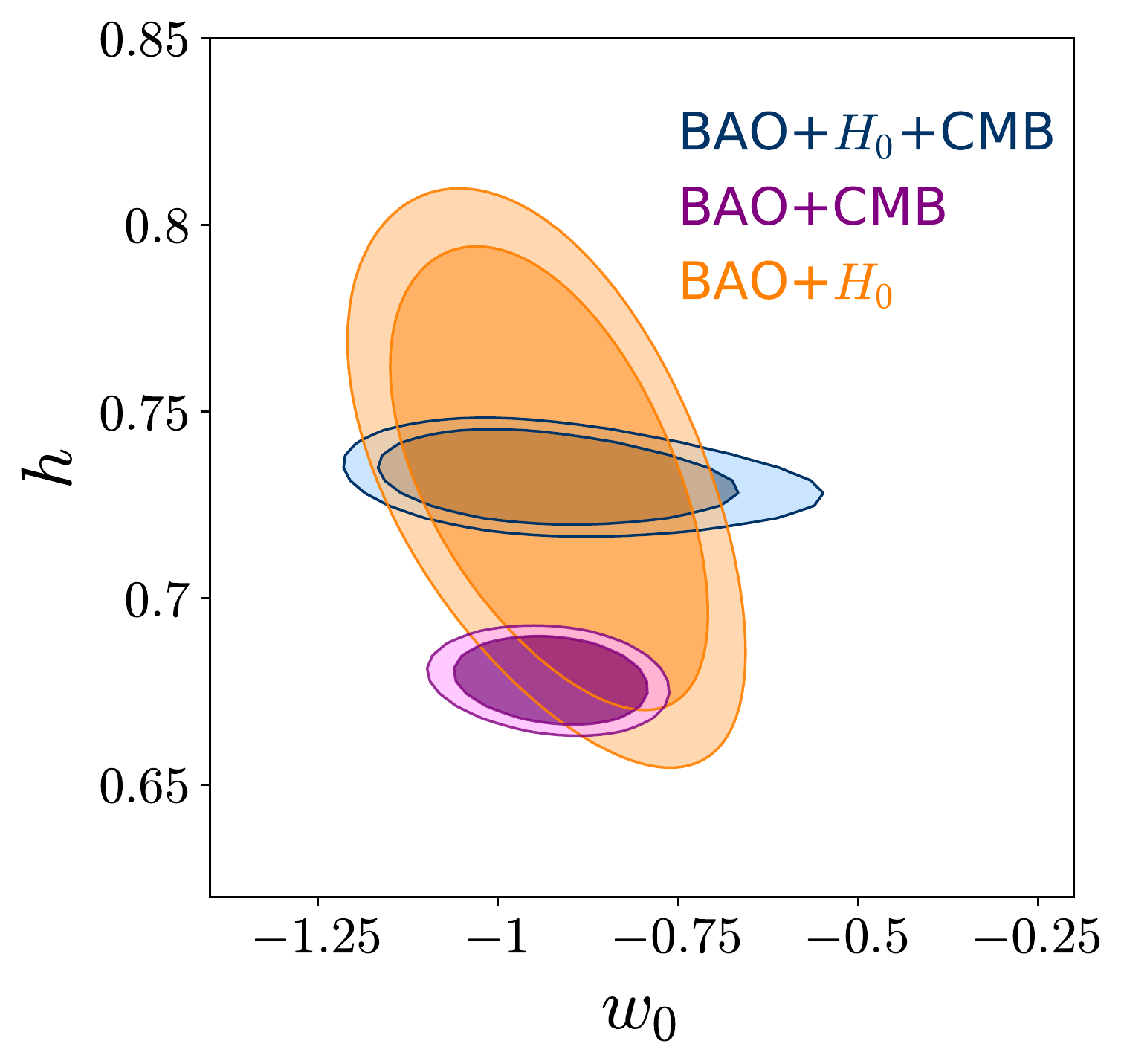}
	\end{subfigure}\quad
	\caption{Contour plots displaying the $1\sigma$ and $2\sigma$ confidence levels for the data sets presented in Table \ref{table:results} on $w_0$-$w_i$ (left), $w_0$-$\Omega_M$ (center), and $w_0$-$h$ space (right). }
	\label{fig:contours1}
\end{figure*}

Results from this section are  discussed below and summarized in table \ref{table:results}, and in figures \ref{fig:bfveos}-\ref{fig:contours1}. 

Local measurements (labeled model A in table \ref{table:results})  point to a dynamical DE presenting a very late and abrupt transition ($z_T=0.28$, $q=9.97$) from an initial value $w_i =-0.99$ to a present value $w_0=-0.91$. This behavior is portrayed in figure \ref{fig:bfveos} and corresponds to the black dot-dashed curve. The value for $H_0$ holds in agreement with the reported measurement from AR16 used as prior for this calculation.

The dynamics for DE resulting from the use of BAO data and CMB reduced likelihood  (outcome B, Table \ref{table:results})  indicates the  preference for a steep transition ($q=9.8$)  from the  initial value $w_i=-0.77$ to the present value $w_0=-0.92$  at a pivotal redshift $z_T=0.63$.  This corresponds to the  dotted line in figure \ref{fig:bfveos}.  The value for $\omega_c$ lies within the range imposed by CMB priors and the BFV for  $H_0$ is lower, in agreement with P15 (\cite{Ade:2015xua}).

Model C  in table \ref{table:results}  shows that  a late time and smooth transition ($z_T$=1.31, $q=1.5$)  was preferred by data, with an initial  value  $w_i=0$ to a present value  $w_0=-0.96$. The blue dashed line in \ref{fig:bfveos} displays this particular dynamics. 
The amount of  matter is very similar  in DE and $\Lambda$CDM models (cases C and $C_\Lambda$), however we obtain a larger   amount of DE
$\rho_{DE} > \rho_\Lambda$ at present time  and therefore  a larger $H_0$.  We see that the  dynamics of DE allows to consistently fit  the variables from
CMB along with the local value of $H_0$,  since the inclusion of $H_0$ in model C only increased $\chi^2$ by 0.2\%  compared to model B.
However,  from Table \ref{table:resultsLambda} we see that the addition of $H_0$ in $\Lambda$CDM model ($B_{\Lambda}$ and $C_{\Lambda}$) severely penalizes the fit by increasing $\chi^2$  by $19\%$, showing a tension  in the value for $H_0$ from  CMB and local measurements.

\begin{figure*}
	\centering
	\begin{subfigure}[t]{0.49\textwidth}
		\includegraphics[width=0.9\linewidth]{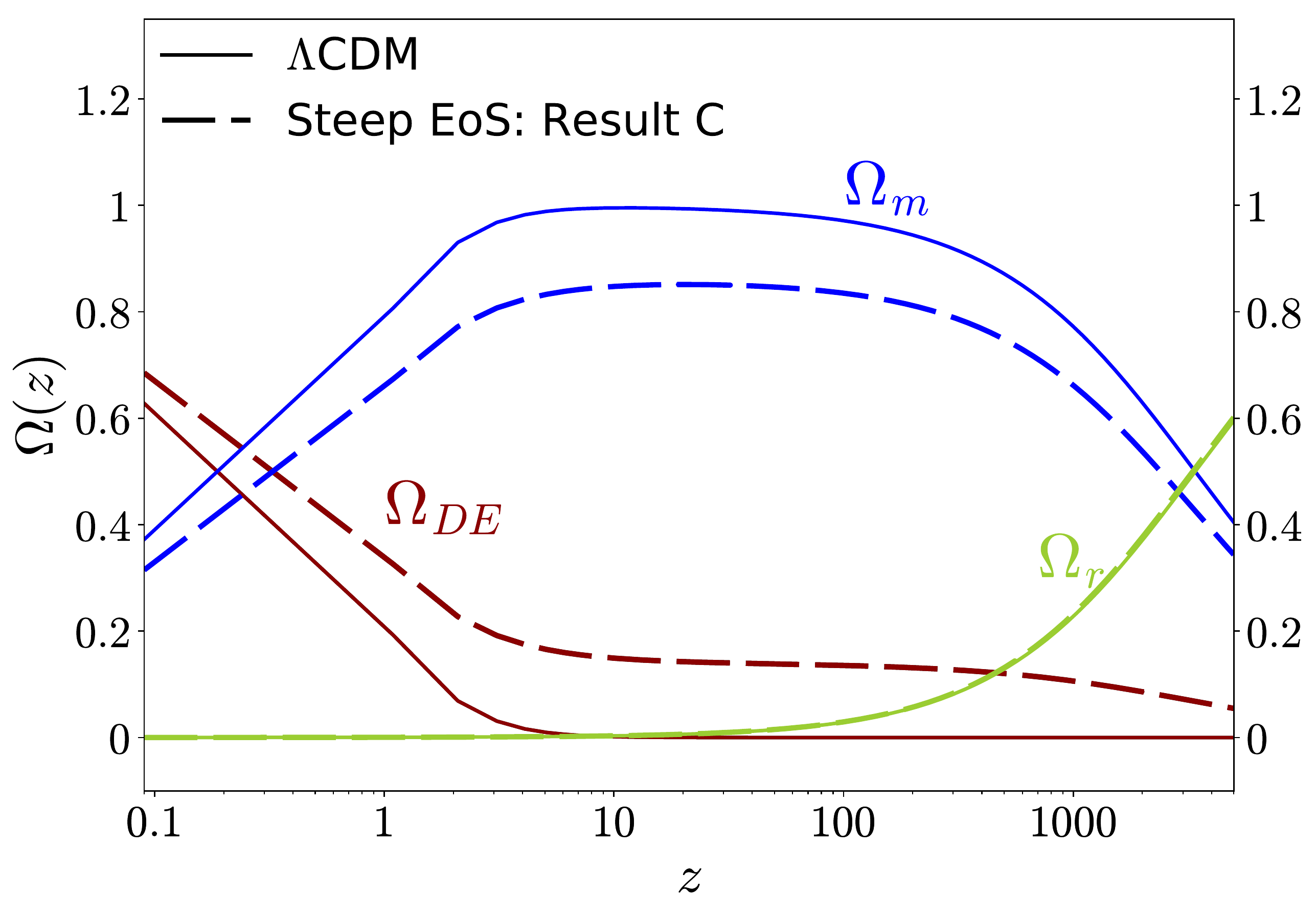}
		\captionsetup{width=0.9\linewidth}
			\caption{Evolution of $\Omega_{DE}(z)$ (red), $\Omega_{m}(z)$(blue) and $\Omega_{r}(z)$(green) for a Cosmological Constant solution (solid line) and for DE (dashed line) analysis.}
			\label{fig:omegaz}
	\end{subfigure}
	\begin{subfigure}[t]{0.49\textwidth}
		\hspace{-0pt}
		\includegraphics[width=0.90\linewidth]{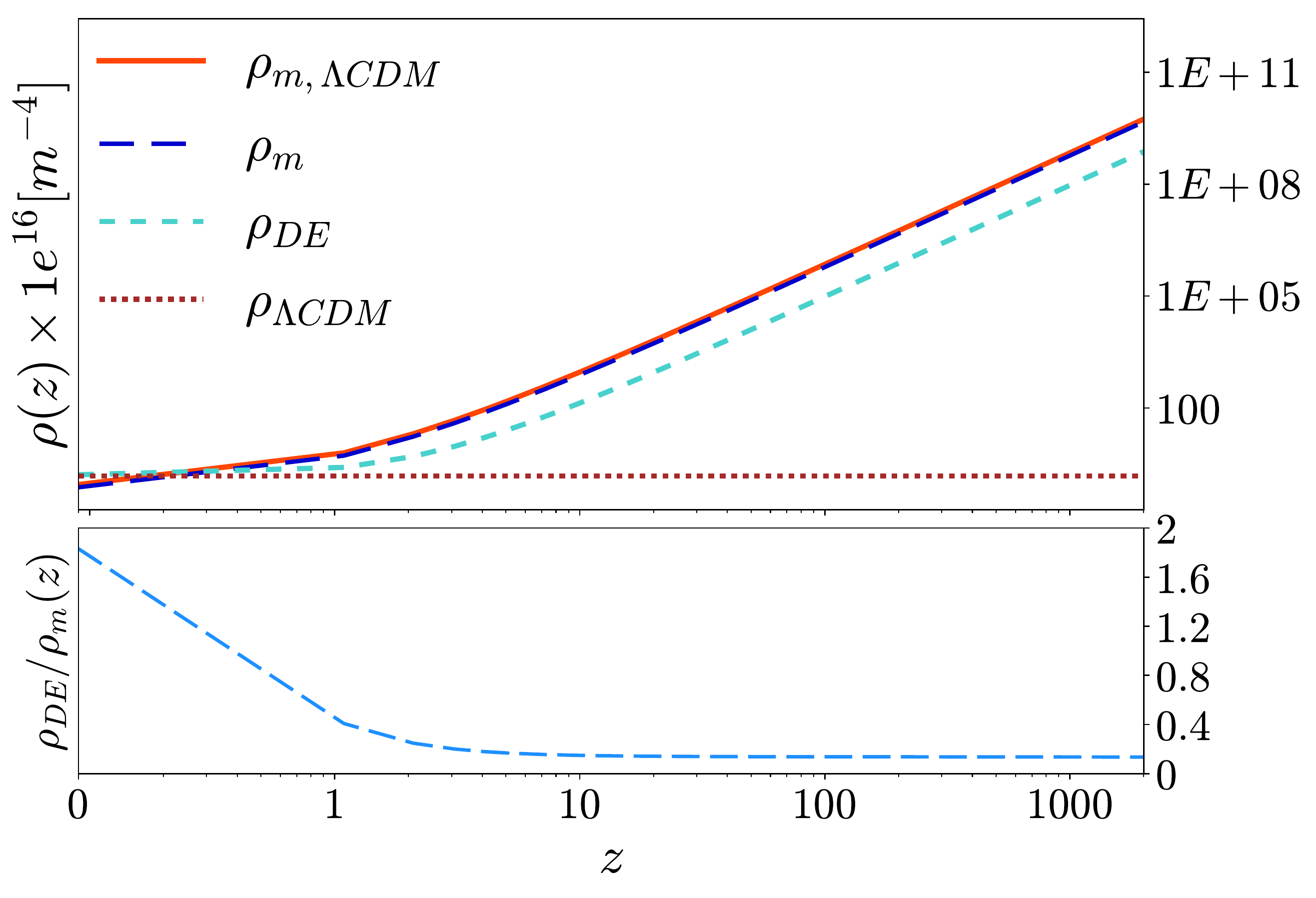}
		\captionsetup{width=\linewidth}
		\caption{Matter density for DE (long-dashed) and for $\La$CDM (solid line), along with DE density (dashed line) and $\rho_{\La}$ (dotted line). 
		Bottom panel: the ratio of DE density to ordinary matter (CDM and baryons)}
			\label{fig:rhos}
	\end{subfigure}
\caption{Comparison of DE following the dynamics resulting from BAO+$H_0$+CMB analysis (result C) and  for $\La$CDM.}
		\label{fig:densities}
\end{figure*}

In this case, the DE density at early times is not negligible since it has $w_i=0$. Figure \ref{fig:densities} shows that its contribution at decoupling is of order   $\Omega_{DE}=10\%$, adding  an extra component that behaves like dust ($\propto a^{-3}$) at large redshifts.  The ratio of DE density to ordinary matter ($\omega_c+\omega_b$) is nearly constant from  $z\gtrapprox5$ and has a value $\rho_{DE}(z_*)/\rho_m(z_*)=0.16$ (Figure \ref{fig:rhos}).
This changes several cosmological parameters, for instance the equivalence epoch, $a_{eq}\equiv \rho_r(a_{eq})/\rho_m(a_{eq})$, is smaller modifying the distance to the last scattering surface  and the sound horizon at recombination. This is an interesting toy model worthwhile of further studies, and it will also impact  CMB power spectrum and Large Scale Structure formation.

Having a non-negligible DE at earlier times, allows to put better constraints on its parameters: $\{w_i, q, z_T\}$.

A  DE component which is non-negligible at early times as been studied in the literature and is known as Early Dark Energy (see for example \cite{Linder:2008nq}).

From both, table \ref{table:results} and figure \ref{fig:contours1} we can draw the following general results. 
The value for $w_0$ is tightly constrained by observations.
The scenario $w_0 = -1$ is included within 1$\sigma$ error for all  the cases.
Ge\-ne\-ra\-lly speaking, for the outcomes where DE density becomes negligible at earlier times, we obtained weak or no constraints for the initial value of the EoS, $w_i$, the transition time, $z_T$, and the exponent $q$. In all outcomes, the values $q=1$ and $z_T=1$ are contained within 1$\sigma$ of significance
Figure \ref{fig:contours1}  shows that $w_i$ is highly degenerated with  $w_0$. 
The results for $\La$CDM are summarized in table \ref{table:resultsLambda}.

\begin{table*}	
	\begin{center}
		\begin{tabular}{|c|c|c|c|c|c|c|c|c|c|c|c|}
			\hline 
			\multicolumn{11}{|c|}{Steep Equation of State for DE}\\
			\hline \hline
			Alias & Data sets used & $\chi^2$  & $w_0$ & $w_i$& $q$  &$z_T$  & $\omega_c$ & $H_0$ & $\Om$ & $\rho_{DE}$ \\

			\hline \hline			
	
			A & BAO + $H_0$ & 9.59   &$-0.92^{+0.15}_{-0.14}$ & $-0.99$($\leq$-0.67) & 9.97& 0.28 &   0.1568$^{+0.0244}_{-0.0208}$& $73.22^{+4.2}_{-4.1}$  & $0.334^{+0.052}_{-0.044}$ &0.3570\\ 
			
			B & BAO+CMB & 9.77 & $-0.92\pm0.10$   & -0.77 ($\leq -0.27$)&  9.8 & 0.63($\geq 0.10$) & 0.1195$\pm$0.0031 &  67.80$\pm$0.9 &0.308$\pm 0.008$&0.3181\\ 
			
			C & BAO+CMB+$H_0$ & 9.79 & $-0.96^{+0.22}_{-0.17}$ &  $0^{+0.04}_{-0.02}$  & $1.5^{+1.3}_{-0.5}$ & $1.31^{+1.42}_{-0.44}$ &  0.1195$\pm$0.0034 & $73.26\pm1.0$ & $0.264\pm0.008$ &0.3950\\			
			\hline
				\end{tabular}
		\caption{BFV and 1$\sigma$ errors for the free parameters  as result  from the combined analysis of BAO data (table \ref{table:datarbao}) along with the local value of $H_0$ \eqref{H0} and CMB priors \eqref{eq:covmatcmb}. The value for $\Om$ and $\rho_{DE}\equiv\Omega_{DE}h^2$ were derived as explained in the text. 
	}
		\label{table:results}
	\end{center}
\end{table*}	
%

\begin{table*}	
	\begin{center}
		\begin{tabular}{|c|c|c|c|c|c|c|}
			\hline 
			\multicolumn{7}{|c|}{$\La$CDM}\\
			\hline \hline
			Alias & Data sets used & $\chi^2$&$\omega_c$   & $H_0$ &$\Om$ & $\rho_{\La}$ \\
			\hline \hline
			
			\hline
			$A_{\Lambda}$&BAO + $H_0$ & 10.05 &$0.1476\pm 0.0052$ & $73.56^{+2.0}_{-2.3}$ & $0.3139^{+0.023}_{-0.026}$  & 0.3712\\
			
				$B_{\Lambda}$&BAO + CMB & 11.74  & $0.1201^{+0.0088}_{-0.0099}$  & $70.20\pm0.5$& $0.2889\pm0.004$ &0.3504 \\	
			$C_{\Lambda}$&BAO  + CMB + $H_0$& 13.98  & $0.1203\pm0.0017$  & $70.99\pm0.5$& $0.2829\pm0.004 $ &0.3614 \\
			\hline
		\end{tabular}
		\caption{Similar to Table \ref{table:results} but  assuming  $w=-1$  as equation of state. The reported parameters are  $\omega_c$ and $H_0$. The value for $\Omega_{m}$  and $\rho_{\La}\equiv\Omega_{\La}h^2$ were derived as explained in the text.}
		\label{table:resultsLambda}
	\end{center}
\end{table*}	

\section{GROWTH OF  PERTURBATIONS}
\label{sec:perturbs}

We are interested in studying the effect that the transition featured by the EoS \eqref{eq:eos} has in the evolution of matter overdensities well inside the horizon in the matter-DE domination era. We do so by means of the following system of  linearized equations:

\begin{subequations}
	\label{eq:perturbscoupled}
	\begin{align}
	a^2\delta_m''(a)+&a\frac{3}{2}\left[1-w(a)\Omega_{DE}(a)\right]\delta_m' (a) \label{eq:Mcoupled}\\ \nonumber
	- &\frac{3}{2} \left[ \Omega_m(a) \delta_m(a) + \Omega_{DE}(a) \delta_{DE}(a) \right] = 0   \\ \nonumber	
	a^2\delta_{DE}''(a)+&a\frac{3}{2}\left[1-w(a)\Omega_{DE}(a)\right]\delta_{DE}'(a)  \\ 	
	\nonumber
	+& \left(\frac{c_s^2 k^2}{a^2 H^2(a)} - \frac{3}{2} \Omega_{DE}(a)\right)\delta_{DE}(a) \\ 
	\label{eq:DEcoupled}
	-& \frac{3}{2} \Omega_{m}(a) \delta_m(a) =  0,
	\end{align}
\end{subequations}
where $\delta_m\equiv\frac{\delta\rho_m}{\rho_m}$ and $\delta_{DE}\equiv\frac{\delta\rho_{DE}}{\rho_{DE}}$ represent the matter and DE density contrasts, respectively; $w(a)$ is the equation of state \eqref{eq:eos} as function of scale factor, $\Omega_{DE}$ and $\Om$ are the DE and matter fractional densities, $H(a)$, the Hubble function and $k$ the Fourier wave number. 

The term $c_s^2$  in equation \eqref{eq:DEcoupled}  represents the speed of sound for the DE. We can split it as the sum of an adiabatic contribution and an \emph{effective} or \emph{non-adiabatic} part:
\begin{equation}
\label{eq:cs2}
c_s^2\equiv\frac{\delta p}{\delta\rho}= c^2_{ad}+c^2_{eff}.
\end{equation}

$c^2_{ad}$ stands for the adiabatic speed of sound, defined as
\begin{equation}
\label{eq:c2ad}
c^2_{ad} \equiv \frac{dP}{d\rho} = w(a) - \frac{a w'(a)}{3(1+w(a))}
\end{equation} 
which depends solely on the equation of state for the DE fluid and its derivative, whereas $c^2_{eff}$ represents the non-adiabatic contribution to the speed of sound. We model it as a constant and take some values, $c^2_{eff}= 0, 1/3, 1$.

\begin{figure}  
	\centering
	\includegraphics[width=\linewidth]{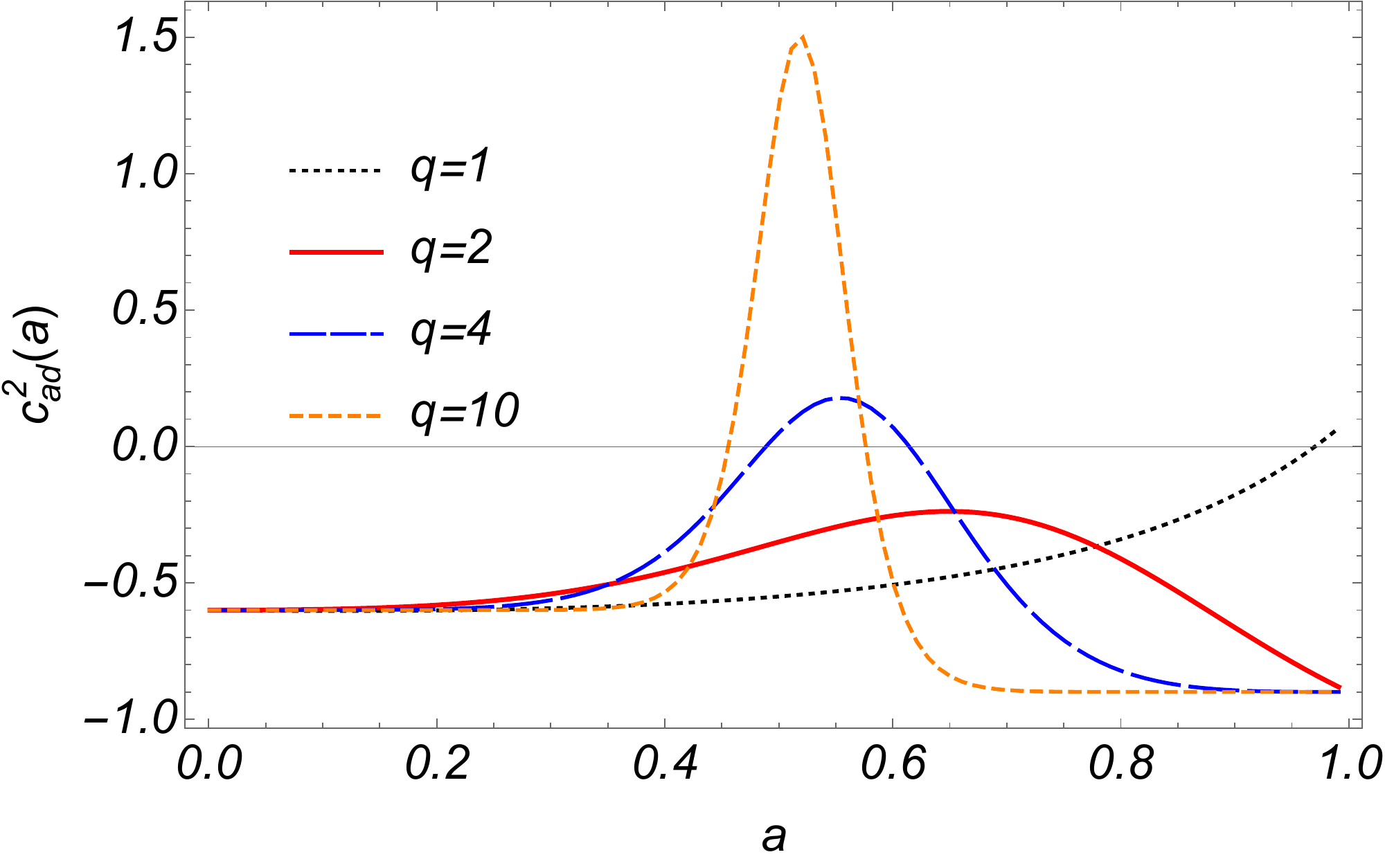}
	\caption{Adiabatic sound speed $c^2_{ad}$ as function of scale factor assuming the EoS in equation \eqref{eq:eos} with $w_0=-0.9$, $w_i=-0.6$, $z_T=1$ and different values for the exponent: $q=1$ (dotted black curve), $q=2$ (dot-dashed red), $q=4$ (dashed blue) and $q=10$ (solid orange line). }
	\label{fig:soundspeed}
\end{figure}

Figure \ref{fig:soundspeed} shows the effect a transition in the EoS has in $c^2_{ad}$. We note that the greater the value for $q$, the sharper the bump in the adiabatic speed of sound. For this part of the analysis we take the values for $\{ w_0, w_i, q, z_T\}$ as those resulting from  A in Table \ref{table:results}.

Adiabatic initial conditions were assumed and the value for $\delta_{m}$ was fixed to be $\delta_m(a_{ini})=10^{-5}$ at the time of entrance for the k-mode into the horizon, whose value was chosen to be $k=0.01Mpc^{-1}$.

\begin{figure}
	\centering
	\includegraphics[width=\linewidth]{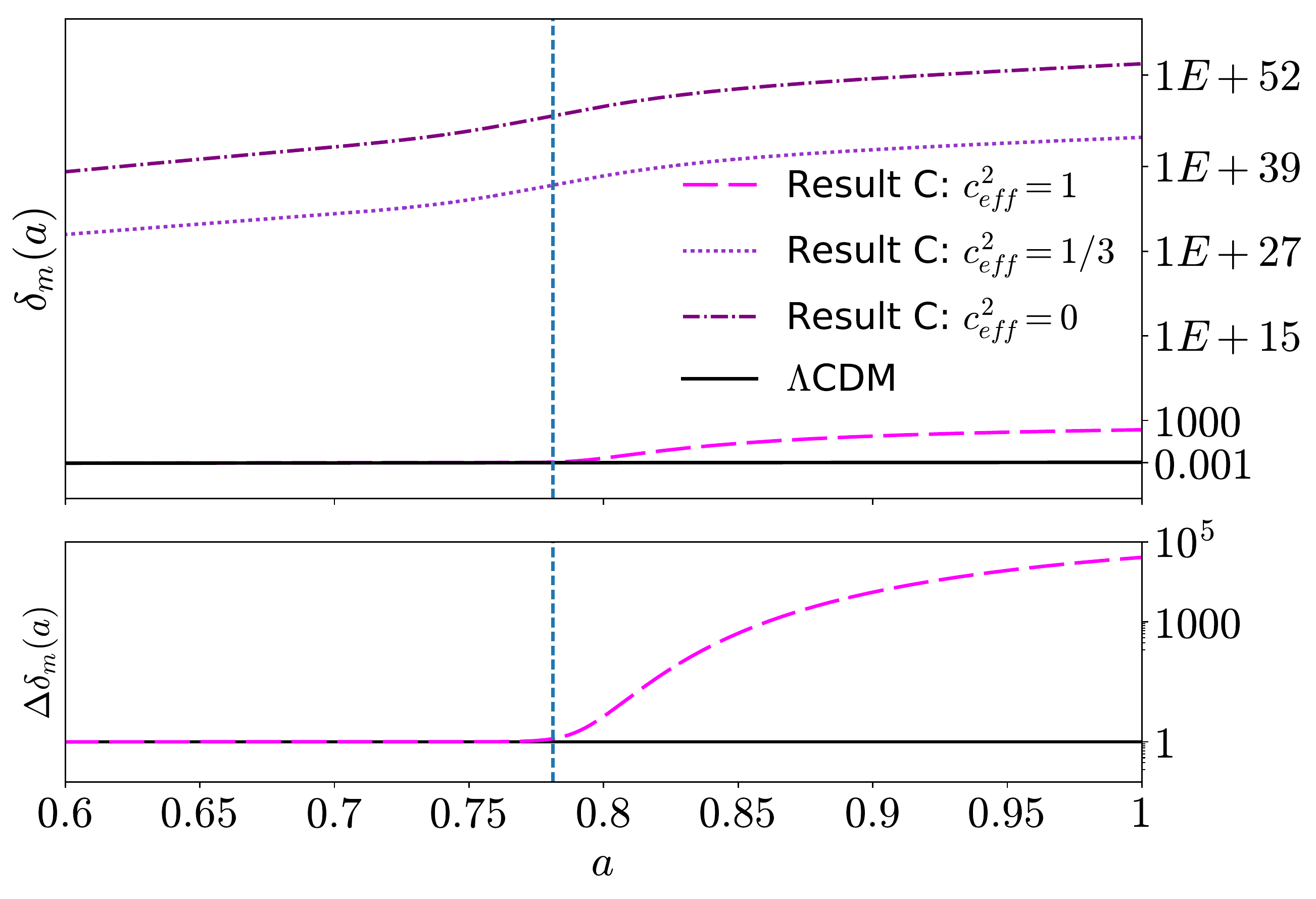}	
	\caption{(Upper panel) Solution for $\delta_m$ from the system of equations \eqref{eq:perturbscoupled} taking values $c_{eff}=1$  (dashed pink line), $c_{eff}=1/3$ (violet dotted line),  and $c_{eff}=0$(purple dot-dashed line) added to the adiabatic speed of sound, $c^2_{ad}$ \eqref{eq:c2ad}. The dashed vertical line marks the transition time, $a_T$.
	(Lower panel) Ratio of solution with $c^2_{eff}=1$ to $\La$CDM, $\Delta\delta_{m}\equiv\frac{\delta_{m}(a)}{\delta_{m,\La CDM}}$.}
	\label{fig:effectivecs}	
\end{figure}

Figure \ref{fig:effectivecs} displays the solution $\delta_{m}$ from equation \eqref{eq:perturbscoupled} for an DE EoS modeled by result A (table \ref{table:results}) and $c^2_{eff}=0, \frac{1}{3}, 1$. We note that the effect of $c^2_{eff}$ is to reduce the magnitude in the growth of $\delta_{m}$, keeping the shape the same. It is also notorious that the evolution of $\delta_{m}$ becomes very non-linear when solved coupled to $\delta_{DE}$, due to the term $\left(\frac{c_s^2 k^2}{a^2 H^2(a)} - \frac{3}{2} \Omega_{DE}(a)\right)\delta_{DE}(a)$ in equation \eqref{eq:DEcoupled}. 

\begin{figure}
	\centering
	\includegraphics[width=\linewidth]{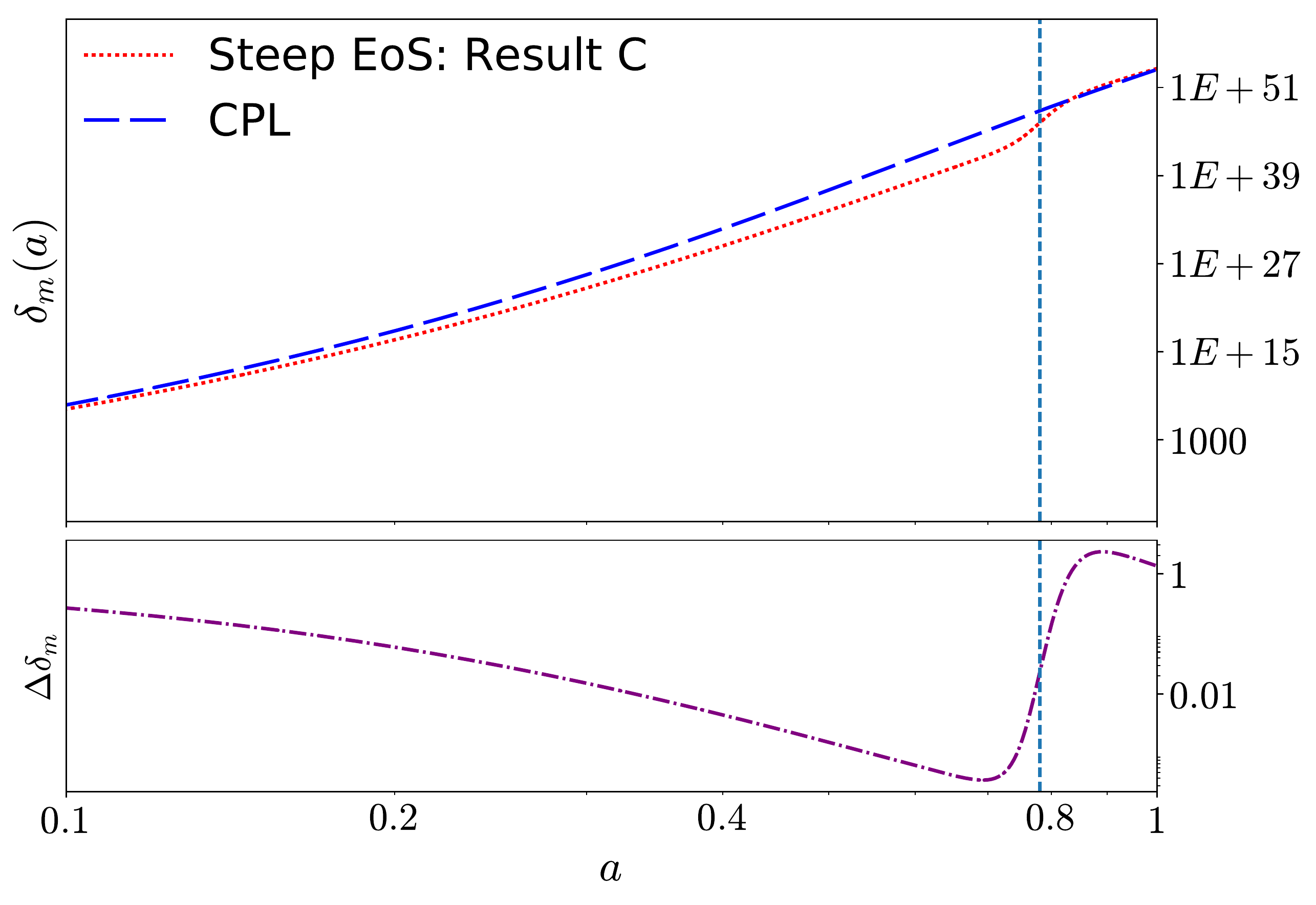}	
	\caption{(Upper panel)Growth of matter overdensities from the system of equations \eqref{eq:perturbscoupled} taking $c^2_{eff}=0$ for the ``BAO + $H_0$" model  and its corresponding CPL  limit, \emph{i. e.}, $q = z_T = 1$.
		(Lower panel) Ratio  of ``BAO + $H_0$" solution to CPL limit, $\Delta\delta_{m}\equiv\frac{\delta_{m}(a)}{\delta_{m,CPL}}$}
	\label{fig:dmatoCPL}.
\end{figure}

\begin{figure}
	\centering
	\includegraphics[width=\linewidth]{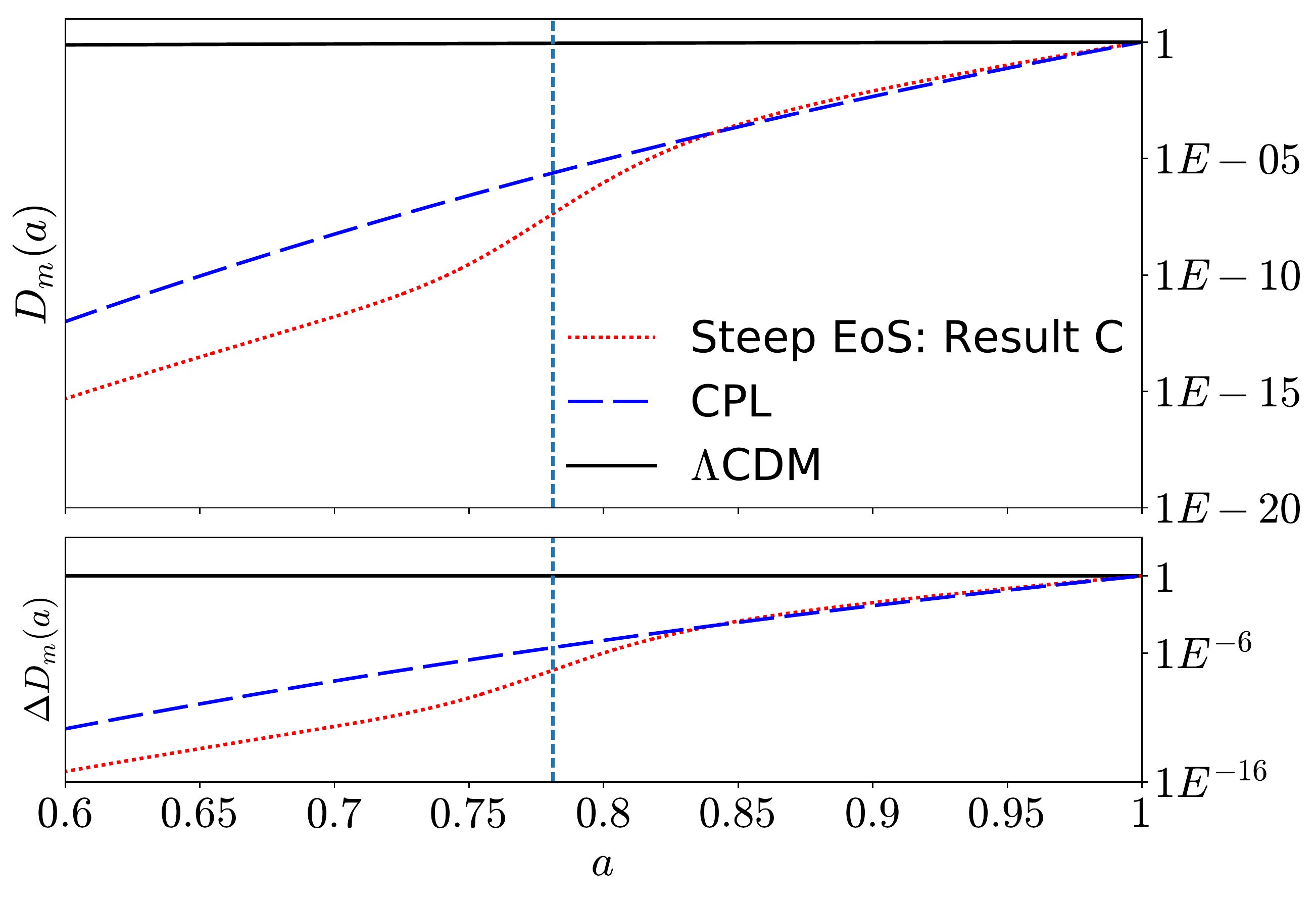}
	\caption{(Upper panel)Same as in figure \ref{fig:dmatoCPL} but displaying the growth function normalized to the present day. 
		(Lower panel) Ratio of ``BAO + $H_0$" and CPL solutions to $\La$CDM scenario, $\Delta D_{m}\equiv\frac{D_{m}}{D_{m,LCDM}}$.}
	\label{fig:coupledperturbs}
\end{figure}

The transition performed in the model  A occurs at $z_T=0.28$ with an steepness given by $q=9.97$. The corresponding time of transition, $a_T=1/(1+z_T) = 0.78$,  is marked by a blue dashed vertical line in figures \ref{fig:effectivecs}, \ref{fig:dmatoCPL}, and \ref{fig:coupledperturbs}.

In figure \ref{fig:dmatoCPL} we analyze in more detail the effect of a steep transition. We fix $c^2_{eff}=0$ to focus on the effect of $c^2_{ad}$ only (displayed in figure \ref{fig:soundspeed}).  For comparison we take the CPL with the same values for $w_0$ and $w_i$ as in Result A and we take the ratio of both solutions. The result is displayed in the lower panel of figure \ref{fig:dmatoCPL}. We note the difference between $q=1$ and $q=9.8$ at the transition time, $a_T$, as a sudden increase during the transition.

Figure \ref{fig:coupledperturbs} shows the normalized growth function $D_m(a)\equiv\frac{\delta_{m}(a)}{\delta_m(a_0)}$ to the present value for the same models as in figure \ref{fig:dmatoCPL}. The bottom panel shows the ratio to $\La$CDM instead.

It is customary to take  $\delta_{DE}=0$. In such case, the system of equations \eqref{eq:perturbscoupled} reduces to:
\begin{equation}
a^2\delta_m''+a\frac{3}{2}\left(1-w(a)\Omega_{DE}(a)\right)\delta_m'- \frac{3}{2}\Omega_m(a)\delta_m=0
\label{eq:perturbsback}
\end{equation}

\begin{figure}
	\centering
	\begin{subfigure}[t]{0.5\textwidth}	
		\captionsetup{width=5cm}
		\includegraphics[width=\textwidth]{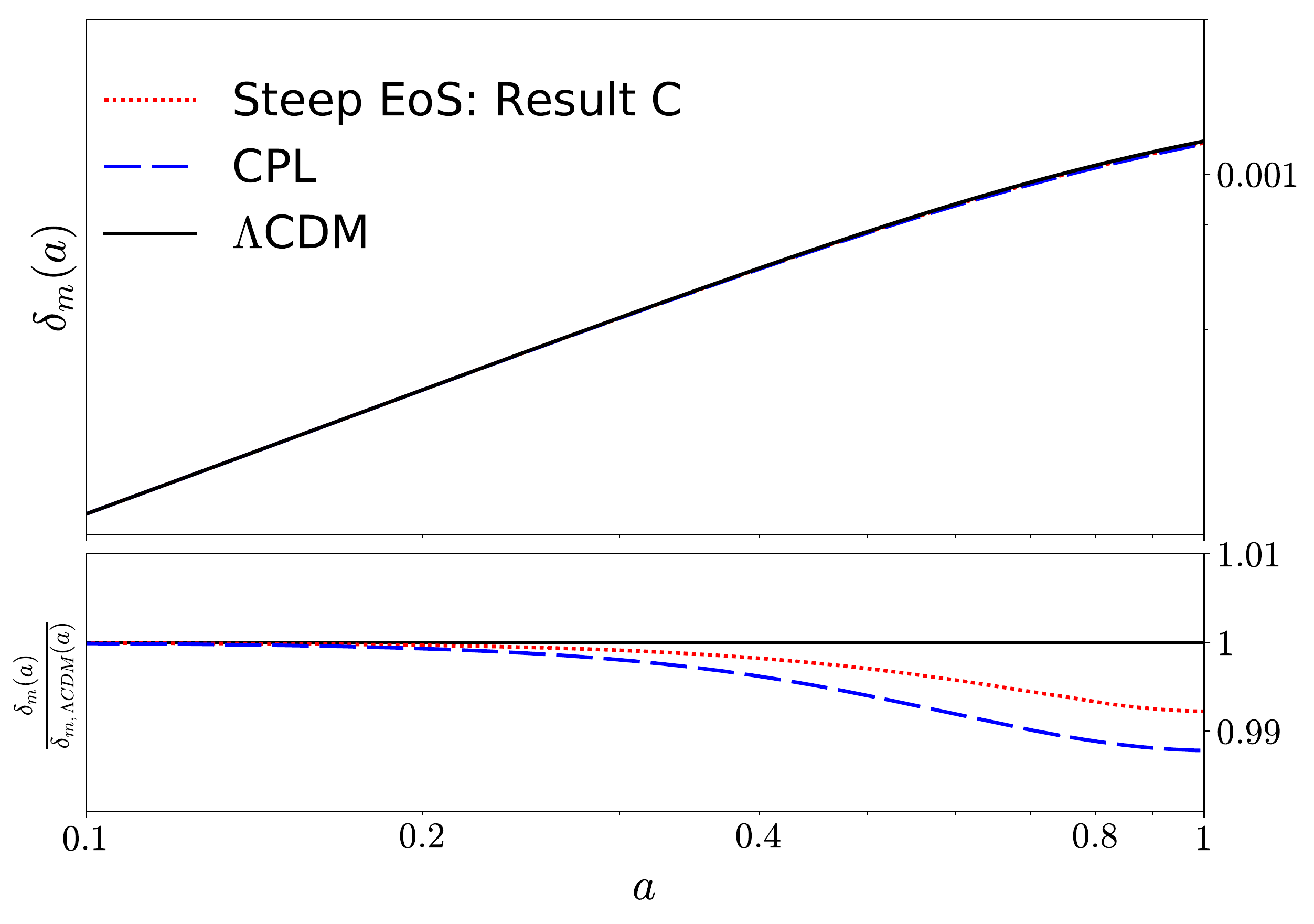}
	\end{subfigure}\quad
	\caption{(Upper panel)Growth of matter overdensity taking $\delta_{DE}=0$ and the model ``BAO+$H_0$" as DE EoS. The CPL limit, (dashed blue line) and  $\La$CDM (solid black line) are also shown. 
	(Lower panel) Ratio of CLP and ``BAO+$H_0$" solutions to $\La$CDM: $\delta_m(a)/\delta_{m,LCDM}(a)$.}
	\label{fig:backgroundperts}
\end{figure}

In figure \ref{fig:backgroundperts} we show the solution to equation \eqref{eq:perturbsback} for the case of a $\La$CDM scenario, the best fit corresponding to Result A in table \ref{table:results},  and its  CPL limit (\ref{eq:CPL}).
The bottom panel  shows the relative difference from each model to $\Lambda$CDM, this is $ \delta_{m}(z))/\delta_{m,\Lambda}(z)$, where $\delta_{m,\Lambda}(z)$ co\-rres\-ponds to the growth of matter contrast when we assume a Cosmological Constant solution.
From  $ \delta_{m}(z))/\delta_{m,\Lambda}(z)$ we find  differences from $\Lambda$CDM of around 1$\%$ for a $k$-mode of $k=0.01Mpc^{-1}$.

 A deeper analysis on the impact of DE perturbations will be subject of a next paper \cite{Jaber2016perturbs}.

\section{SUMMARY AND CONCLUSIONS}

\label{sec:Conclusions}

We presented a parametrization for the EoS of  DE and found the constraints deduced by using  BAO measurements contained in Table \ref{table:datarbao} combined with the latest local determination of Hubble constant (\cite{localhubble}). Additionally we used the compressed CMB likelihood from Planck (\cite{Ade:2015xua}), by means of the sound horizon at decoupling, $\theta_*$, and $\omega_ch^2$ .

The constraints for the free parameters, $\{w_0, w_i, q, z_t, \omega_c, H_0\}$, and their 68$\%$ errors resulting from the combined analysis of the datasets were obtained.

Our results show that a dynamical DE is favored by data and that a steep transition is preferred by local measurements, \emph{i.e.} BAO and $H_0$, and by BAO with CMB Planck observables (figure \ref{fig:bfveos}).

Whereas for a  $\Lambda$CDM model, the tension between the local determination of $H_0$ (\cite{localhubble}) and the value derived from Planck (\cite{Ade:2015xua}) remains (table \ref{table:resultsLambda}), we find that it is possible to simultaneously conciliate the observations from BAO, $H_0$ and CMB in a single model (table \ref{table:results}) by means of  a dynamical Dark Energy.

For the perturbative analysis it was  shown  that the feature from the shape of $c_{ad}$ due to the particular form of the  EoS got imprinted in the evolution of $\delta_m$. We modeled the speed of sound  splitting it into an adiabatic contribution and an effective term, $c_s^2 = c^2_{ad} + c^2_{eff}$, where $c^2_{eff}$ encapsulates the physics beyond the EoS of the dark fluid.  The addition of this effective term did not erase the features from the bump in the adiabatic speed of sound but suppresses the exponential evolution of the over-densities by several orders of magnitude.  This should be studied in detail and will be subject to discussion in a future paper  \cite{Jaber2016perturbs}.

The solution $\delta_{DE}=0$ is only exact for the case of a cosmological constant. To be consistent we need to take DE perturbations into account. The solution to \eqref{eq:perturbscoupled} with $c_s^2$ taken as the adiabatic contribution \eqref{eq:c2ad} for the model ``BAO+$H_0$" reported in table \ref{table:results} was shown in figure \ref{fig:coupledperturbs}. From this we saw that if we take $c_s^2 = c_{ad}$, the solutions were highly unstable and became non-linear extremely fast during the evolution of over-densities. 

To summarize, the study of dynamics of Dark Energy is a matter of profounds implications for our understanding of the Universe and its physical laws.  Although the measurements from CMB  are  the most precise data sets in Cosmology, the best way to analyze the properties  of  DE comes from the low redshift regime, where the BAO feature is the most robust cosmic ruler. In this work  we have contributed towards that direction,  and we have presented the constraints for a dynamical DE model coming from the analysis of BAO distance measurements combined with the most recent $H_0$ determination and CMB information.

\acknowledgments
We acknowledge financial support from  DGAPA-PAPIIT IN101415  and Conacyt Fronteras de la Ciencia 281 projects. MJ thanks Conacyt for financial support and to the attendees of MACSS for helpful discussions.

\appendix

\section{Appendix}
\label{appendix}
For the  BAO measurements we use the $\chi^2$ function defined as
\begin{equation}
\label{eq:chi2bao}
\chi^2_{BAO} = \vec{y}_{BAO}^T \mathcal{C}_{BAO}^{-1}\vec{y}_{BAO},
\end{equation}
where $\vec{y}_{BAO}\equiv\rbao^{Th}(\vec{\alpha}| z_i) -\rbao^{Obs}(z_i)$ is the difference between theoretical prediction for $\rbao(z)$ according to \eqref{eq:rbao} and the values listed in table \ref{table:datarbao}, and $\mathcal{C}^{-1}_{BAO}$ is the inverse of the covariance matrix containing the observational errors for the measurements. 
Since the data points used in this work are not correlated we have a diagonal matrix whose elements are the square-root of the errors reported in table \ref{table:datarbao}.

Additionally to BAO data we make use of the local determination of $H_0$  by means of
\begin{equation}
\label{eq:chi2h0}
\chi^2_{H_0} = \frac{\left[H(\vec{\alpha}|z=0)-H_0^{Obs}\right]^2}{\sigma_{H_0}^2}
\end{equation}
where $H(\vec{\alpha}|z=0)$ is the Hubble function \eqref{eq:Hz} evaluated in $z=0$ taking the model described by $\vec{\alpha}$ and $H_0^{Obs} = 73.24 km \cdot s^{-1} Mpc^{-1}$ and $\sigma_{H_0}=1.74 km \cdot s^{-1} Mpc^{-1}$, according to the primary fit obtained by A.~Riess {\it et al.} in \cite{localhubble}.

Finally, for the CMB, we use the determination of $\theta_*$ and  $\omega_c$, made by  the Planck Collaboration (P.~A.~R.~Ade {\it et al.} 2015, \cite{Ade:2015xua}). In particular we use \emph{Planck TT + TE + EE + low P} values from the table 4 of \cite{Ade:2015xua} with the covariance matrix displayed in \eqref{eq:covmatcmb}.
With this matrix we can build the $\chi^2_{CMB}$ function:
\begin{equation}
\label{eq:chi2cmb}
\chi^2_{CMB} =  \vec{y}_{CMB}^T \;\; \mathcal{C}_{CMB}^{-1} \;\; \vec{y}_{CMB}
\end{equation}

where $\vec{y}_{CMB}$ is the corresponding data vector defined as $\vec{y}_{CMB} = 
\begin{bmatrix}
\omega_c^{Th} - \omega_c^{Planck}, & \theta_*(\vec{\alpha})^{Th} - \theta_*^{Planck}
\end{bmatrix}^{T}$, and $\mathcal{C}_{CMB}^{-1}$ is the inverse of matrix \eqref{eq:covmatcmb}.

\bibliography{SteepDE-Biblio}
\bibliographystyle{unsrt}

\end{document}